\documentclass[A4,10pt]{smart_2017_Paper}
\usepackage{amsmath}
\usepackage{amsfonts}
\usepackage{amssymb}
\usepackage{booktabs,amsthm,amsbsy,latexsym,graphicx,color}
\usepackage{rotating}
\usepackage{subfigure}
\usepackage{gensymb}
\usepackage{upgreek}
\usepackage[round, sort, numbers]{natbib}
\usepackage{bm}
\usepackage{float}
\usepackage{relsize}

\setcitestyle{square}

\title{A Finite Strain Constitutive Model Considering Transformation Induced Plasticity for Shape Memory Alloys under Cyclic Loading}

\author{LEI XU$^{*}$, THEOCHARIS BAXEVANIS$^{\dag}$ AND DIMITRIS LAGOUDAS$^{*}$$^{\ddag}$}

\heading{Lei Xu, Theocharis Baxevanis and Dimitris Lagoudas}

\address{$^{*}$Department of Aerospace Engineering, Texas A\&M University\\ 
College Station, TX 77840, USA
\and
$^{\dag}$ Department of Mechanical Engineering, University of Houston\\
	Houston, TX 77204, USA
\and
$^{\ddag}$ Department of Material Science and Engineering, Texas A\&M University\\ 
College Station, TX 77840, USA}


\abstract{Many engineering applications of Shape Memory Alloys (SMAs) involve passing back and forth through phase transformation many times. Repeated phase transformation develops permanent deformations originating from the significant distortion that phase transformation induces at the austenite--martensite interfaces and grain boundaries. This distortion drives dislocation activity resulting in an observable macroscopic Transformation-Induced Plastic (TRIP) deformation, which occurs at effective stress levels much lower than the plastic yield limit of the material. TRIP strains may accumulate up to 20\% during the lifetime of an SMA component and thus a finite strain constitutive model is required for simulating their response.  In this work, a 3-D finite strain model is developed based on logarithmic strain which is the only strain measure whose rate is equal to the stretching, that describes phase transformation and TRIP deformation in a thermodynamically consistent setting. The model is implemented in Abaqus finite element software through a user defined material subroutine (UMAT). Boundary value problems such as strip and a torque tube under both pseudoelastic and actuation cyclic loadings are performed to test the capabilities of the newly proposed model. }

\begin{document}

\section{INTRODUCTION}

SMAs have the ability to recover a pre-defined shape when subjected to appropriate thermo-mechanical loading. Since the discovery of the shape memory effect, SMAs has been extensively investigated to be used as sensors, controllers and actuators etc. towards building smart system integrated with adaptive and intelligent functions.

Over the several past decades, a substantial number of SMA constitutive models at the continuum level have been proposed, most of them being based on the infinitesimal strain assumption within the small deformation regime.  A review of such models can be found in Boyd and Lagoudas~\cite{boyd1996}, Birman and November~\cite{birman1997}, Raniecki and Lexcellent~\cite{raniecki1992,raniecki1998}, Patoor et al.~\cite{patoor2006,patoor1996}, Hackl and Heinen~\cite{hackl2008}, Levitas and Preston~\cite{levitas1998,levitas2002}. The majority of above mentioned models are dealing with stable material response that exhibits no evolutionary features. However, many engineering applications of SMAs involve passing back and forth through phase transformation many times. Repeated phase transformation develops permanent deformations originating from the significant distortion that phase transformation induces at the austenite martensite interfaces and grain boundaries especially for the initial cyclic loadings. This distortion drives dislocation activity resulting in an observable macroscopic transformation-induced plastic deformation, which occurs at effective stress levels much lower than the plastic yield limit of the material. From experiment results, TRIP strains may accumulate up to 20\% during the lifetime of an SMA component, and thus a finite strain constitutive model accounting large strains and rotations is required for give better predications.

A three-dimensional constitutive model at finite strain regime for SMAs incorporating TRIP is proposed in this work. The adopted logarithmic strain measure and rate~\cite{xiao1997,xiao1997hypo,xiao2006,xu2017finite} enables the model to effectively capture the material response under finite strains with rotations, and attribute modeling advantages in the presence of large accumulated plastic strain under cyclic thermo-mechanical loading for SMAs. The martensitic volume fraction, transformation strain, and TRIP strain are used to characterize material nonlinear behavior while back stress tensor and a drag stress term are used to introduce the effects of the kinematic and isotropic hardening in the model.

To this end, the paper is organized as follows. In Section~\ref{Preliminary}, we represent some preliminaries on kinematics in continuum mechanics. Section~\ref{Model} will concentrate on the thermodynamic framework to formulate the constitutive model by using logarithmic strain and logarithmic rate. Boundary value problems will be addressed to test the capability of proposed model in Section~\ref{Result}. At the end, we summarize this paper with conclusion in Section~\ref{conc}.

\section{PRELIMINARY} \label{Preliminary}
\subsection{Kinematics}
Let $ \mathbf{X} $ and  $ \mathbf{x}=\chi(\mathbf{X},t) $  denote the position vector of the material points of a body $\mathcal{B} $ in the reference (undeformed) configuration at initial time $ t_{0} $ and at the current (deformed) configuration at time $ t $, respectively. The deformation process from the initial configuration to current configuration can be characterized by the deformation gradient,
\begin{equation}\label{Deformation}
\mathbf{F}(\mathbf{x},t) =\frac{\partial \mathbf{x}}{ \partial \mathbf{X}}  
\end{equation}
Then, the velocity gradient $\mathbf{L}$ is defined through as follows:
\begin{equation}\label{Vel_g}
\mathbf{L} =\mathbf{\dot{F}}\mathbf{F} ^{-1} 
\end{equation}
The velocity gradient $\mathbf{L}$ can be additively decomposed into a symmetric part called the rate of deformation tensor, i.e. $\mathbf{D}$, plus an anti-symmetric part called the spin tensor, i.e. $\mathbf{W}$,
\begin{equation}\label{S_P_tensor}
\begin{array}{c}
\mathbf{L = D + W}; \quad
\begin{cases} \mathbf{D} =\dfrac{1}{2} \mathbf{(L+L^{T})}, \vspace{5pt}  \\ 
\mathbf{W} =\dfrac{1}{2} \mathbf{(L-L^{T})}, \\  
\end{cases}
\end{array}
\end{equation}
The following polar decomposition formula is well known, in which $\mathbf{R}$ is the rotation tensor and $\mathbf{V}$ is the left stretch.
\begin{equation}\label{Polar_Dec}
\mathbf{F = VR}
\end{equation}
The left Cauchy-Green tensor $\mathbf{B}$ is defined by
\begin{equation}\label{LCG_tensor}
\mathbf{B} = \mathbf{FF}^{T} =\mathbf{V}^2 
\end{equation}
The logarithmic strain of Eulerian type $ \mathbf{h} $ is given through,
\begin{equation}\label{Log_strain}
\mathbf{h} = \frac{1}{2} \ln\mathbf{ {B}} =\mathbf{\ln {V}}
\end{equation}
The Kirchhoff stress tensor is defined as equation \ref{Kirch_stess}, where $J$ is the determinant of the deformation gradient $\mathbf{F}$.
\begin{equation}\label{Kirch_stess}
\mathbf{\bm\uptau} = J \bm{\sigma}
\end{equation}
%

\subsection{Logarithmic rate and Logarithmic spin}\label{Logarithmic}
In finite deformation analysis, the rate of the deformation tensor $\mathbf{D}$ can be additively decomposed into an elastic part $\mathbf{D}^e$ and a dissipative part $\mathbf{D}^{dis}$. One of the main tasks is to adopt an appropriate objective rates to achieve the principle of objectivity in its rate form constitutive equations. Many objective rates, such as the Jaumann, Green--Naghdi, and Truesdell rates,  have been proposed by different researchers. However, none of them was able to set up a direct relation between the rate of deformation tensor $ \mathbf{D} $ and an objective rate of strain measure, thus many spurious phenomena, such as shear stress oscillation, dissipative energy or residual stress accumulated in elastic deformation, are observed when these rates are being used. Until recently the so-called logarithmic rate proposed by Xiao et al.~\cite{xiao1997,xiao1997hypo,xiao2006}, Bruhns et al.~\cite{bruhns1999self,bruhns2001large,bruhns2001self}, Meyers et al.~\cite{meyers2003elastic,meyers2006choice} successfully resolved such self-inconsistent issues. It was shown that the logarithmic rate of the logarithmic strain $\mathbf{h}$ of Eulerian type is identical with the rate of deformation tensor $\mathbf{D}$, which is expressed as:
\begin{equation}\label{eq:Log_strain_rate}
\mathring{\mathbf{h}}^{log} = \dot{\mathbf{h}}+\mathbf{h} \mathbf{ \Omega}^{log}-\mathbf{ \Omega}^{log}\mathbf{h}= \mathbf{D}
\end{equation}
Where $\mathring{\mathbf{h}}^{log}$ means the logarithmic rate of logarithmic strain and $\dot{\mathbf{h}}$ is the conventional time rate of logarithmic strain.  $ \mathbf{ \Omega}^{log} $ is called logarithmic spin introduced by Xiao and Bruhns \cite{xiao1997} with its explicit expression as:
\begin{equation}\label{eq:Log_spin}
\mathbf{\Omega}^{log} = \mathbf{W}+ \sum_{i \neq j}^{n}  \big(\frac{1+(\lambda_{i}/\lambda_{j})}{1-(\lambda_{i}/\lambda_{j})}+\frac{2}{•\ln (\lambda_{i}/\lambda_{j})}\big) \mathbf{B}_i \mathbf{D} \mathbf{B}_j
\end{equation}
Where $ \mathbf{W} $ is the spin tensor; $\lambda_{i,j} (i,j=1,2,3...) $ are the eigenvalues of left Cauchy-Green tensors $ \mathbf{B} $; $ \mathbf{B}_{i,j} $ are the corresponding subordinate eigenprojections of $ \mathbf{B}$.


\subsection{Additive decomposition of logarithmic strain }\label{AdditiveStrain}
Starting from the additive decomposition of the rate of deformation tensor $\mathbf D$,
\begin{equation}\label{eq:add_D}
\mathbf{D}=\mathbf{D}^{e}+\mathbf{D}^{tr}+\mathbf{D}^{p}
\end{equation}
The total stress power supplied from outside working on body $ \mathcal{B} $ per unit volume can be calculated and additively decomposed into,
\begin{equation}\label{eq:power}
P_{stress\_power}=\bm{\uptau}:\mathbf{D}=\bm{\uptau}:\mathbf{D}^{e}+\bm{\uptau}:\mathbf{D}^{tr}+\bm{\uptau}:\mathbf{D}^{p}
\end{equation}
From energy point of view, additive decomposition of kinematics in finite deformation analysis can be interpreted as total stress power being split into a recoverable part of $\bm{\uptau}:\mathbf{D}^{e}$, an irrecoverable dissipation part  $\bm{\uptau}:\mathbf{D}^{tr}$ for transformation process and a plastic dissipation part $\bm{\uptau}:\mathbf{D}^{p}$ associated with plasticity deformation.

By virtue of equation \ref{eq:Log_strain_rate}, elastic part $\mathbf{D}^{e}$, transformation part  $\mathbf{D}^{tr}$  and plastic part $\mathbf{D}^{p}$ in equation \ref{eq:add_D} can be rewritten as the logarithmic rate of elastic logarithmic strain ${\mathbf{h}}^{e}$, logarithmic rate of transformation logarithmic  strain ${\mathbf{h}}^{t}$ and  logarithmic rate of plastic logarithmic  strain ${\mathbf{h}}^{p}$ respectively,
\begin{equation}\label{eq:add_h_rate1}
\mathring{\mathbf{h}}^{e\_log}=\mathbf{D}^{e};~~\mathring{\mathbf{h}}^{t\_log}=\mathbf{D}^{tr};~~\mathring{\mathbf{h}}^{p\_log}=\mathbf{D}^{p}
\end{equation}
Combine equation \ref{eq:add_D} and equation \ref{eq:add_h_rate1}, the following equation can be obtained.
\begin{equation}\label{eq:add_h_rate2}
\mathring{\mathbf{h}}^{log}=\mathring{\mathbf{h}}^{e\_log}+\mathring{\mathbf{h}}^{t\_log}+\mathring{\mathbf{h}}^{p\_log}
\end{equation}
Apply logarithmic corotational integration\cite{khan1995continuum} in equation \ref{eq:add_h_rate2} on both sides, the following additive decomposition of total logarithmic strain can be received. Namely, the total logarithmic strain can be additively split into an elastic part, a transformation part and a plasticity part. 
\begin{equation}\label{eq:add_h}
\mathbf{h}=\mathbf{h}^{e}+\mathbf{h}^{t}+\mathbf{h}^{p}
\end{equation}


\subsection{FINITE STAIN SMA MODEL WITH TRIP }\label{Model}
\subsubsection{Thermodynamic potential of constitutive model}
A Gibbs free energy expression is proposed~\cite{Bo1999_TRIP_3, lagoudas2004TRIP} 
\begin{equation}\label{GIBBS_explicit} 
\begin{aligned}
G=  -\dfrac{1}{2 \rho_{0}} \bm{\uptau} : \bm{S}\bm{\uptau} - \dfrac{1}{\rho_{0}}  \bm{\uptau} :[~\bm{\alpha}(T-T_0)+\mathbf{h}^t+\mathbf{h}^p]  -\frac{1}{\rho_{0}} \int^{\xi}_0 (\bm{\beta}:\frac{\partial\mathbf{h}^t}{\partial\xi}+\eta)d\xi        \\+c \Big[(T-T_0)-T\ln (\dfrac{T}{T_0}) \Big]-s_0(T-T_0)+u_0
\end{aligned}
\end{equation}
where $ \bm{\uptau} $ is the Kirchhoff stress tensor, $ T $ is the temperature and $\mathbf{\Upsilon}=\{\mathbf{h}^t,\mathbf{h}^p,\bm{\beta},\eta,\xi\}$ is the set of internal state variables, with $\bm{\beta}$, $\eta$, $ \xi $  denoting a back-stress tensor, a drag-stress scalar, and the martensite volume fraction, respectively. $\bm{\beta}$ and $\eta$ describe hardening effects. Moreover, $ \bm{S}(\xi)$ is the fourth-order compliance tensor by using rule of mixtures, $  \bm{\alpha}$ is the second order thermoelastic expansion tensor, $ c $ is effective specific heat, $ s_0, u_0 $ are effective specific entropy at reference state and effective specific internal energy at reference state, respectively, and $ T_0 $ is the reference temperature.

Following the standard Coleman-Noll procedure in dissipation inequality, the following constitutive relations can be obtained
\begin{equation}\label{h_Cons_f}
\mathbf h =  -\rho_{0}\frac{\partial G}{\partial \bm {\uptau}} = \bm{S}\bm{\uptau} + \bm\alpha:(T-T_0)+ \mathbf{h}^t+ \mathbf{h}^p
\end{equation}   
\begin{equation}\label{entropy_Cons_f}
s =  -\rho_{0}\frac{\partial G}{\partial T} = \frac{1}{\rho_0}  \bm{\uptau} : \bm\alpha  +c \ln (\dfrac{T}{T_0}) -s_0
\end{equation}
where  $s$ denotes the entropy.
Apart from the constitutive relation obtained, the following strict form of the dissipation inequality can also be derived
\begin{equation}\label{Dissipation_State_V2}
-\rho_{0}  \frac{\partial G}{\partial \mathbf{h}^t} :\mathring{\mathbf{h}^t} -\rho_{0}  \frac{\partial G}{\partial \mathbf{h}^p} :\mathring{\mathbf{h}^p} -\rho_{0}  \frac{\partial G}{\partial \xi}\dot{\xi} \geqslant 0
\end{equation}
The following section will concentrate on setting up the evolution equations between different internal state variables.


\subsubsection{Evolution equation for transformation strain}\label{Evolution_Trans}

One key assumption of the classical SMAs model postulated by Boyd and lagoudas~\cite{boyd1996} is that any change in the current microstructural state of the material is strictly a result of a change in the martensitic volume fraction. We thus propose the following evolution relationship between $\xi$ and  $ \mathbf{h}^t $ .
\begin{equation}\label{Trans_Evol}
\begin{aligned}
{\mathring{\bm h}}^{t}= \bm{\Lambda}  \dot{\xi},  \ \  \bm\Lambda=\begin{cases}\bm{\Lambda}_{fwd}, \; \dot{\xi}>0, \vspace{5pt} \\ \bm{\Lambda}_{rev}, \; \dot{\xi}<0, \end{cases}\\
\end{aligned}
\end{equation}
where, $\bm\Lambda_{fwd}$ is the transformation direction tensor during forward transformation process, while $\bm\Lambda_{rev}$ is the transformation direction tensor during reverse transformation process, they are defined respectively in equation \ref{Evo_tr}. It is worth to point out that the rate in equation \ref{Trans_Evol} is logarithmic rate for finite deformation analysis defined in Section \ref{Logarithmic}
\begin{equation}\label{Evo_tr}
\bm\Lambda_{fwd}=
\frac{3}{2} H^{cur} 
\frac{\bm{\uptau}^{\text{eff}'}}{\bar{\uptau}^{\text{eff}}}; \ \  \ \     
\bm\Lambda_{rev}=
H^{cur} \frac{\mathbf h^{t}}{\bar{\mathbf h}^{t}}.
\end{equation}

$ \bm{\uptau}^{\text{eff}} $ is the effective stress tensor, which is defined as equation \ref{S_eff}, the back stress tensor $\bm\beta$ will be introduced shortly. 
\begin{equation}\label{S_eff}
\bm{\uptau}^{\text{eff}}=\bm{\uptau}+\bm\beta
\end{equation}
$ \bm{\uptau}^{'} $ is the deviatoric stress tensor part defined by {\small $ \bm{\uptau}^{'} =\bm{\uptau} -{\small \frac{1}{3}}\textrm{tr}(\bm{\uptau})\mathbf{I} $}, in which $ \mathbf{I} $ is the second order identity tensor. The von mises equivalent stress is given by
\begin{equation}
\bar{\uptau} ={\small \sqrt{{{\small \frac{3}{2}}\bm\uptau^{'}}:\bm{\uptau}^{'}}} 
\end{equation}
$H^{cur}$ denotes the current maximum transformation strain for current stress state. From experiment observation, the maximum transformation strain  $H^{cur}$ is not a constant value throughout all stress state, it is a exponential function depending on current stress state as defined through an curve fitting in equation \ref{Hcur}, in which $H^{max}$ is the maximum saturated transformation strain and $\text{k}_t$ is a curve fitting material parameter.
\begin{equation}\label{Hcur}
H^{cur}(\bm\uptau)= H^{max}(1-e^{-\text{k}_t { \bar{\bm\uptau}^{\text{eff}}}})
\end{equation}
%


\subsubsection{Evolution equation for plastic strain}\label{Evolution_Plastic}
In order to model the large irrecoverable transformation induced plastic strain, a evolution equation for the plastic strain is also proposed here. It has a similar structure as the evolution equation proposed by Lagoudas and Entchev \cite{lagoudas2004TRIP}. Instead of using the Cauchy stress, infinitesimal strain tensor and the conventional time rate for the rate form evolution equation, we used the  Kirchhoff stress, logarithmic strain tensor and the logarithmic rate to reformulate the evolution relation in equation \ref{Evol_plastic}. 
\begin{equation}\label{Evol_plastic}
\begin{aligned}
{\mathring{\bm h}}^{p}= \bm{\Lambda}^p  \dot{\xi},  \ \  \bm\Lambda^p=\begin{cases}\bm{\Lambda}^p_{fwd}, \; \dot{\xi}>0, \vspace{5pt} \\ \bm{\Lambda}^p_{rev}, \; \dot{\xi}<0, \end{cases}\\
\end{aligned}
\end{equation}

In the above rate form evolution equation,  $\bm\Lambda^p_{fwd}$ is the forward TRIP  direction tensor while  $\bm\Lambda^p_{rev}$ is the reverse TRIP direction tensor. They have the explicit expression in equation \ref{Dir_plastic} respectively. The rate here is logarithmic rate.

\begin{equation}\label{Dir_plastic}
\bm\Lambda^p_{fwd}=\frac{3}{2} C^p_1 \dfrac{H^{cur}(\bm\uptau)}{H^{max}}\dfrac{{ \bm{\uptau}^{\text{eff}'}}} {{{\bar{\bm{\uptau}}}^{\text{eff}}} } e^{-\frac{\zeta^d}{C_2^p}} ;\ \ 
\bm\Lambda^p_{rev}=- C^p_1 \dfrac{H^{cur}(\bm\uptau)}{H^{max}}\dfrac{\mathbf h^{t-r}}{\bar{\mathbf h}^{t-r}} e^{-\frac{\zeta^d}{C_2^p}} .
\end{equation}

In the above defined TRIP direction tensor, the accumulated detwinned martensitic volume fraction $\zeta^d$ is introduced through the following equation, the way it was introduced here is based on the assumption that self-accommodating martensitic phase transformation plays no role on the development of the plastic strains \cite{lagoudas2004TRIP}.
\begin{equation}
\zeta^d = \int_0^t |\dot{\xi}^d(t)|dt
\end{equation}
The detwinned martensitic volume fraction $\xi^d$ is introduced as equation \ref{xi^d}
\begin{equation}\label{xi^d}
\xi^d = \dfrac{H^{cur}(\bm\uptau)}{H^{max}}\xi
\end{equation}
%

\subsubsection{Drag stress and back stress}\label{Evolution_back}

In this model, a second order back stress tensor and a scalar drag stress term are introduced to accounting for the kinematic and isotropic hardening features during the transformation process. They are essentially representing microstructure effects such as grain boundaries, misfit between austenite-martensite interface, micro voids and cracks etc. The back stress tensor is defined in a similar way by Lagoudas and Entchev \cite{lagoudas2004TRIP}, which has the tensorial equation \ref{backstress}. The difference is here we used logarithmic strain measure instead of infinitesimal strain measure. 
\begin{equation}\label{backstress}
\bm{\beta} = - \dfrac{\bm{h}^t}{\bar{\bm{h}}^t}  \big[ D^b_1(H^{cur}\xi)+D^b_2(H^{cur}\xi)^{2}+D^b_3(H^{cur}\xi)^{3}
+D^b_4(H^{cur}\xi)^{4}+D^b_5(H^{cur}\xi)^{5} \big]
\end{equation}
The scalar drag stress term is the same as it was in paper Lagoudas and Entchev \cite{lagoudas2004TRIP}.  It is revisited here in equation \ref{dragstress} for clarity .
\begin{equation}\label{dragstress}
{\eta} = - D_1^d \big[-\ln {(1-\xi)} \big]^{\frac{1}{m_1}}+D_2^d\xi 
\end{equation}
%


\subsubsection{Transformation function}\label{Trans_Func}
In this section, a criterion for the occurrence of both the forward and reverse transformation will be defined to make the model completed. After the introduction of evolution equation for internal state variables, we substitute the evolution equation \ref{Trans_Evol} and \ref{Evol_plastic} into the strict form dissipation inequality \ref{Dissipation_State_V2}, the following inequality can be derived:
\begin{equation}\label{Dissipation_xi}
(\bm\uptau:\bm\Lambda-\rho_{0}  \frac{\partial G}{\partial \xi})\dot{\xi}=\pi\dot{\xi}\geqslant 0 
\end{equation}
In the above equation, the quantity $\pi$ is called the general thermodynamic driving force conjugated to $\xi$. Substitute Gibbs free energy expression equation \ref{GIBBS_explicit} into equation \ref{Dissipation_xi}, the explicit expression of general thermodynamic driving force $ \pi $ is obtained:
\begin{equation}\label{Driving_Force}
\begin{aligned}
\pi=\dfrac{1}{2}\bm\uptau:\mathit{\Delta}\bm{S}\bm\uptau+
\bm\uptau^{\text{eff}}:\bm\Lambda+\bm\uptau:\bm\Lambda^{p}
+\bm\uptau:{\Delta}\mathbf{\alpha}(T-T_0) +\eta\\ 
-\rho_0\Delta c\big[ T-T_0-T\ln(\dfrac{T}{T_0}) \big ] + \rho_0\Delta s_0 (T-M^{0s}) + Y
\end{aligned}
\end{equation}
Where $\mathit{\Delta}\bm{S}, \Delta \mathbf{\alpha}, \Delta c, \Delta s_0, \Delta u_0 $ represent the difference of these material parameters between austenitic and martensitic phase by virtue of the rule of mixture. $M^{0s}$ is introduced as a combination of other parameter such that

\begin{equation}\label{Mos}
M^{0s}= T_0 + \frac{1}{\rho_0\Delta s_0}(Y+\rho_0\Delta u_0)
\end{equation}

Consistent with the framework by Qidwai and Lagoudas \cite{qidwai2000}, we are assuming that whenever the thermodynamic driving force reaches a critical value $Y $ or $ (-Y) $, the forward (reverse) phase transformation will take place. Thus transformation function $\Phi$ is introduced here to working as a criterion as defined by equation (\ref{Transfor_Fun}). 
\begin{equation}\label{Transfor_Fun}
\normalfont{\Phi}=\begin{cases}~~\pi - Y, \; \dot{\xi}>0, \vspace{5pt} \\ -\pi - Y, \; \dot{\xi}<0, \end{cases}\\
\end{equation}
To satisfy the second law of thermodynamics and the principle of maximum dissipation for the internal state variables, some certain constraints have to apply on the model, this constraint can be expressed in terms of so-called Kuhn-Tucker conditions:
\begin{equation}\label{Kuhn-Tucker}
\begin{aligned}
\dot{\xi} \geqslant 0; \quad \Phi(\bm\uptau,T,\xi)= ~~\pi - Y \leqslant 0;  \quad  \Phi\dot{\xi}=0;~~~\bf{(A\Rightarrow M)}\\
\dot{\xi} \leqslant 0; \quad \Phi(\bm\uptau,T,\xi)= -\pi - Y \leqslant 0; \quad   \Phi\dot{\xi}=0;~~~ \bf{(M\Rightarrow A)}
\end{aligned}
\end{equation}


\section{NUMERICAL RESULTS}\label{Result}
In order to demonstrate the capabilities of proposed model, two types of problems are presented here in this section. The first problem analyzed is a strip under unaxial tension loading while the other problem is a torque tube under torsion loading. For both problems, isothermal stress-induced transformation case and isobaric thermal-induced transformation case are performed respectively. The results are obtained through using the commercial finite element software Abaqus combined with user defined material subroutine (UMAT). Material parameters used in all the cases are presented in Table \ref{tab1} from the book by Lagouas et al. \cite{lagoudas2008}.

\begin{table}[t!]	
	\caption{Material parameters for SMA  used for the numerical results \cite{lagoudas2008}}
    \label{tab1}
	\renewcommand{\arraystretch}{1}
	\begin{center}
		\begin{tabular}{lr|lr} \toprule
		Parameter    & Value &Parameter & Value  \\ \midrule
		$E_A$   & 70   [GPa] & $D^1_b$  & 3.40 $\times$ 10 $^{3}$ [MPa]\\
		$E_M$   & 50   [GPa] & $D^2_b$ & -2.23 $\times$ 10 $^{5}$ [MPa]\\
		$\nu_A=\nu_M$& 0.33   & $D^3_b$  & 8.32 $\times$ 10 $^{6}$ [MPa]\\
		$\alpha_A=\alpha_M$ & 2.2 $\times$10$^{-5}$ [K$^{-1}$]& $D^4_b$& -1.50 $\times$ 10 $^{8}$ [MPa]\\
		$ H^{max}$    & 0.05 & $D^5_b$ & 1.03 $\times$ 10 $^{9}$ [MPa]\\
		$k_t$ & 0.0052 [MPa$^{-1}$]& $D^1_d$ & 8.0 [MPa]\\
		$\rho c^A$ & 2.12 [MJ/(m$^{3}$K)]&$D^2_d$  & 1.7 [MPa]\\
		$\rho c^M$ & 2.12 [MJ/(m$^{3}$K)]& $m_1$ & 3.5 \\
		$\rho \Delta s_0 $ & -0.422 [MJ/(m$^{3}$K)]& $C_1^P$ & 3.6 $\times$ 10 $^{-3}$\\
		$M^{0s}$& 311.0 [K]&$C_2^P$& 18.0 \\
		& &$Y$& 6.0 [MJ/m$^{3}$] \\
		\bottomrule
		\end{tabular}
	\end{center}
\end{table}
%

\subsection{Uniaxial isothermal loading case}\label{uniaxial_isothermal}

This boundary value problem describes a SMA strip sample under cyclic isothermal uniaxial tensile load up to a maximum value of 800 MPa. During the loading step, the load is linearly increased from initial value 0 MPa to its maximum value 800 MPa, which activates the forward stress-induced  transformation. After the load reach the peak value, it begins unloading from 800 MPa to 0 MPa, which results in a reverse transformation. The temperature is keeping a constant value of 360 K throughout this whole process. Loading cycle is repeated 50 times. The material stress versus strain response curve for this case is depicted in Figure 1. In order to have a clean picture, only several loading cycles are chosen to show in the figure. It can been seen that there is a substantial amount of irrecoverable plastic strain accumulated during this cyclic loading process. The TRIP strain in the only first cycle is about 0.7\% and saturated with a value close to 6\% after 30 cycles. The material response is stabilized and the plastic strain accumulated in the 50$^{th}$ is almost zero. 
\begin{figure}	\label{Strip_Pseudo1}
	\begin{center}
		\includegraphics[width=0.7\columnwidth]{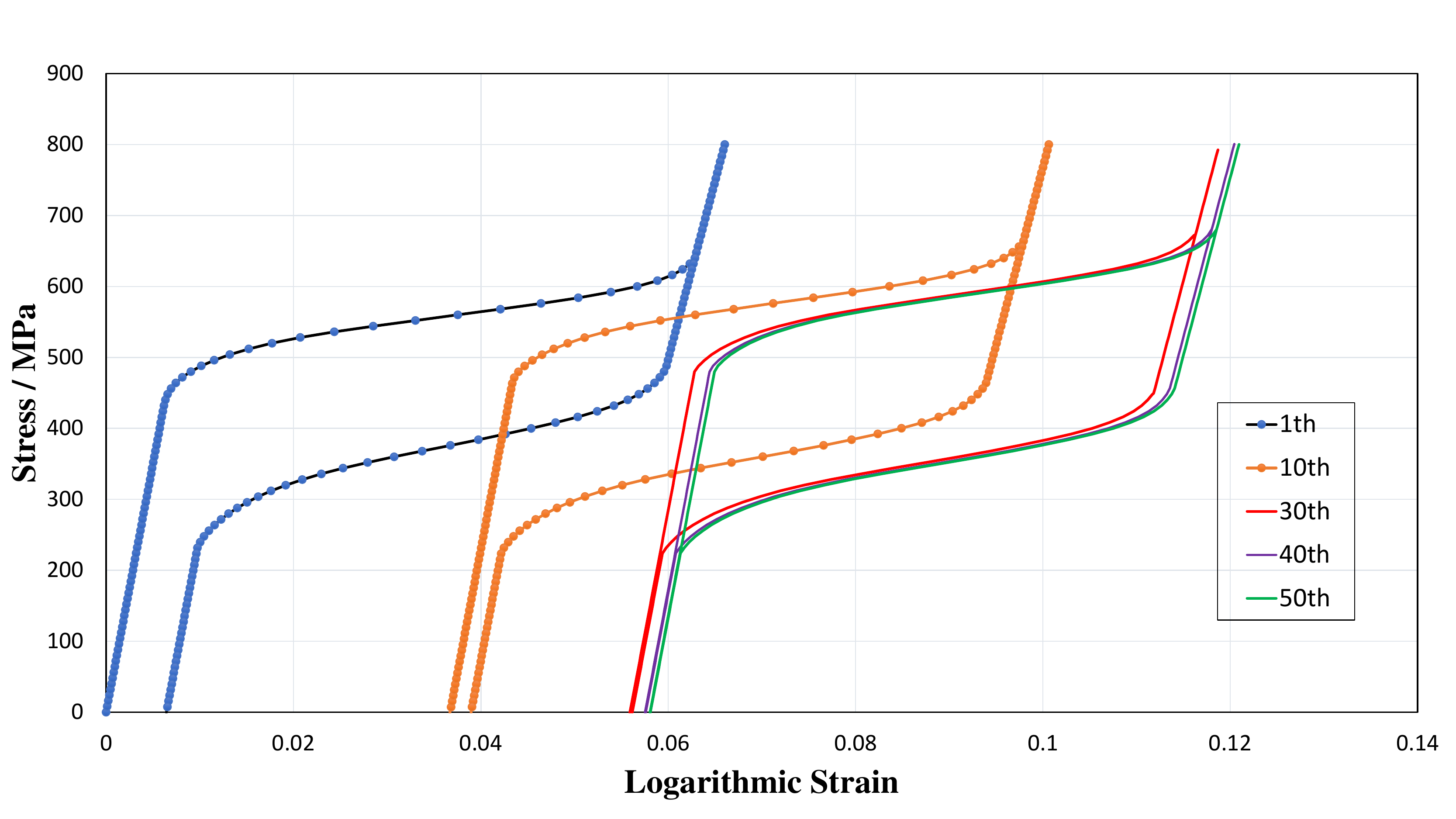}
	\end{center}
	\caption{Pseudoelastic response of strip from 1$^{st}$ to 50$^{th}$ cycle}
\end{figure}

\subsection{Uniaxial isobaric loading case}\label{uniaxial_isobaric}

This boundary value problem is simulating a SMA strip under cyclic actuation loading. The SMA strip is first subjected to a constant uniaxial tensile load of 200 MPa, then it undergoes multiple coolings and heatings by changing the temperature from 360 K to 260 K back and forth. The strain versus temperature curve is plotted in Figure 2, again only several typical curves are chosen to show to have a clean result. As is shown in Figure 2, there is a large amount of irrecoverable plastic strain accumulated during the cyclic actuation loadings. The TRIP strain starts with value around 0.6\% and end up with almost a zero value after 30 cycles, the actuation response is also stabilized and the accumulated TRIP strain saturated with a value around 4.8\% after 50 cycles. \\

\begin{figure}\label {Strip_Actution}
	\begin{center}
		\includegraphics[width=0.7\columnwidth]{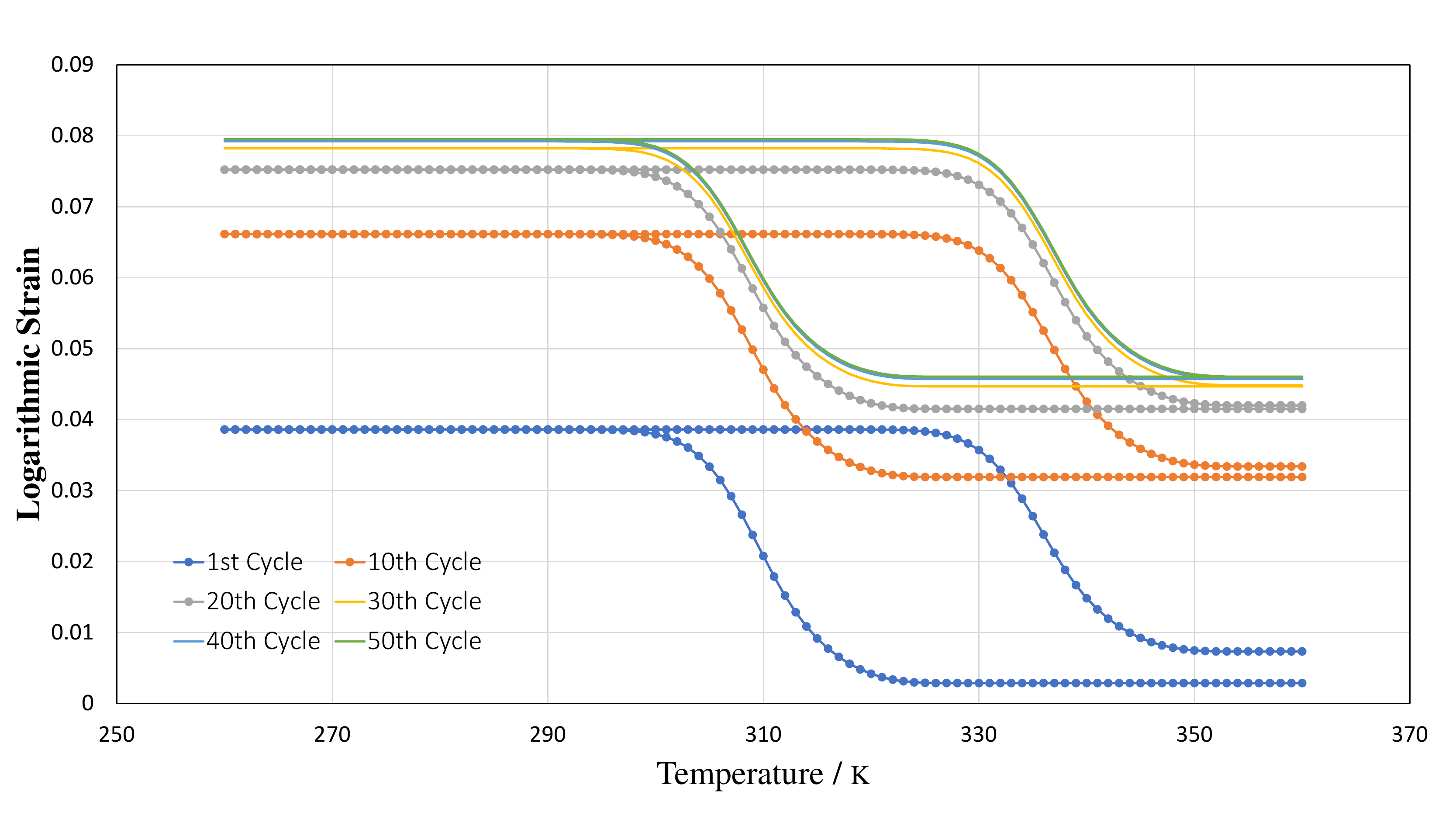}
	\end{center}
	\caption{Actuation response of strip from 1$^{st}$ to 50$^{th}$ cycle}
\end{figure}

\subsection{Torsion  isothermal loading case}
In this section, a torque tube problem is used to demonstrate the capability of proposed model for loading case with rotations. The tube is subjected to a torque linearly changing from its initial 0 value to a maximum value during loading, and linearly decreasing from its maximum value to 0 value for unloading. The temperature is kept as a constant value of 360 $\degree $C throughout the cyclic loading process.The material pseudoelastic stress versus strain curve is plotted in Figure 3. The plotting information comes from a point at the torque tube inner surface. From the figure, it is interesting to see the difference between these results and those obtained from unaxial case in \ref{Strip_Pseudo1}. Apart from the saturated TRIP strain and stabilized material response, the pseudoelastic curve of torque tube shifted down compared to it was in uniaxial case. This can be explained by the internal stress introduced through the training process. Since the shear stress is not uniform along the radial direction, thus the plastic strain accumulated due to phase transformation is also not uniform along radial direction, the outer surface will have larger TRIP strain as a result of the larger shear stress it experienced while the inner part will have a smaller one. A nonuniform shear stress distribution make the response of tube shifted compared to its counterparts in the uniaxial case.  \\

\begin{figure}\label {Tube_Pseudo}
	\begin{center}
		\includegraphics[width=0.7\columnwidth]{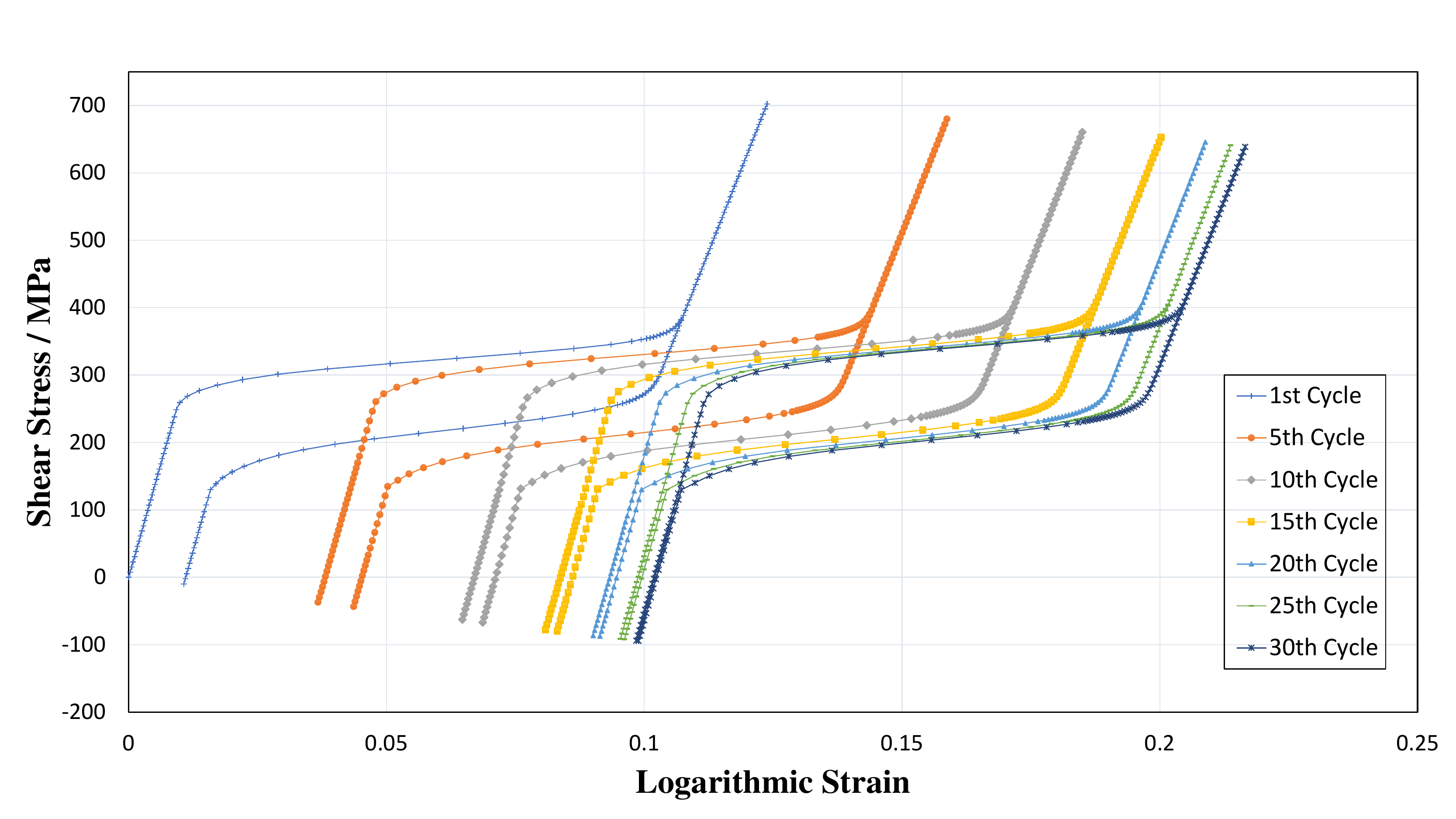}
	\end{center}
	\caption{Pseudoelastic response of torque tube from 1$^{st}$ to 30$^{th}$ cycle}
\end{figure}

\subsection{Torsion  isobaric loading case}
The last problem analyzed is the torque tube under cyclic actuation loading. Similar to the loading case in section \ref{uniaxial_isothermal}, the SMA tube is first subjected to a constant torque, then it will experience the cooling and heatings by changing the temperature from 360 K to 290 K back and forth. The actuation response of torque tube is shown in Figure 4, similar as what observed in Figure 2, there is also a large amount of irrecoverable plastic strain accumulated. The TRIP strain saturates with a value around 9\% in the end, and the actuation response is also stabilized after 30 cycles. 

\begin{figure}\label {Tube_Actuation}
	\begin{center}
		\includegraphics[width=0.7\columnwidth]{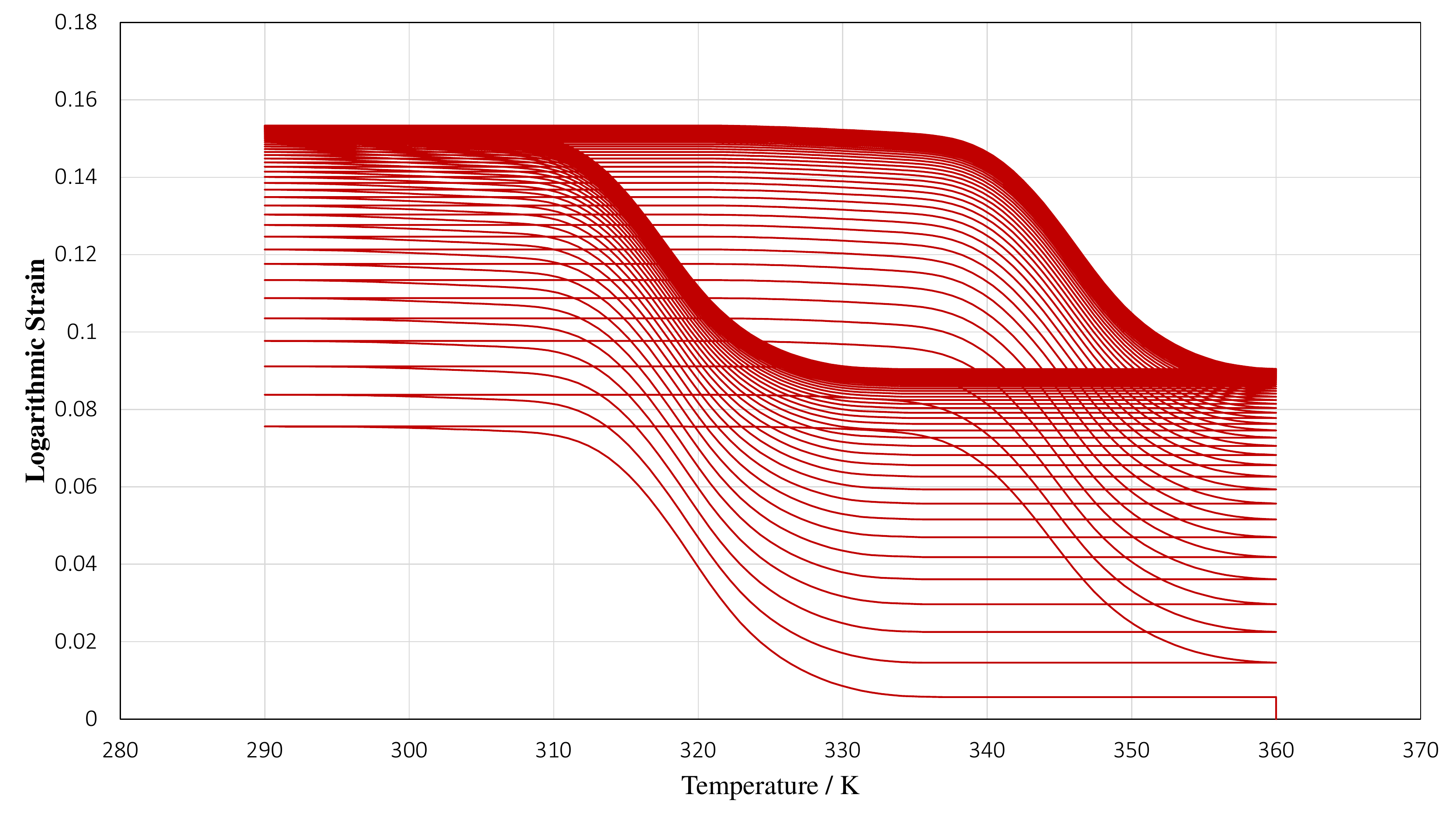}
	\end{center}
	\caption{Actuation response of torque tube from 1$^{st}$ to 30$^{th}$ cycle}
\end{figure}
\section{CONCLUSIONS}\label{conc}
In the paper, a three-dimensional constitutive model at finite strain regime for phase transformation materials with large accumulating transformation induced irrecoverable plastic strain is proposed in this work. Boundary value problems such as a strip and a torque tube under both pseudoelastic and actuation loadings are performed to demonstrate the capabilities of the proposed model. Numerical results of strip problem demonstrated that the proposed model can characterize the material nonlinear evolution features, such as TRIP strain accumulation and its saturation, stabilized material response after certain loading cycles in a good manner. The torque tube problem also showed that the model can predict the internal stress introduced through the material training process. In the future, the presented model will be calibrated through and compared to the experiment results.

\newpage
\bibliographystyle{elsarticle-harv} 
\bibliography{myarticle}

\end{document}